\documentclass[%
 reprint,
showpacs,
amsmath,amssymb,
aps,
pra,
]{revtex4-1}

\usepackage{graphicx}
\usepackage{dcolumn}
\usepackage{bm}

\usepackage{pstricks}

\newcommand{\ket}[1]{|{#1}\rangle}
\newcommand{\bra}[1]{\langle{#1}|}
\newcommand{\prj}[1]{\ket{#1}\bra{#1}}
\newcommand{\ketbra}[2]{\ket{#1}\bra{#2}}
\newcommand{\R}[1]{{\textrm{#1}}}

\newcommand{\cT}{{\mathcal T}}
\newcommand{\cM}{{\mathcal M}}
\newcommand{\cF}{{\mathcal F}}

\newcommand{\Wgeo}[1]{{\mathcal W}_{#1}}

\newcommand{\deltac}{\delta_{{\rm c}2}}

\newcommand{\diffD}{D}

\begin{document}


\title{Cavity cooling of a trapped atom using Electromagnetically-Induced Transparency}

\author{Marc Bienert$^1$}
 \email{marc.bienert@physik.uni-saarland.de}
\author{Giovanna Morigi$^{1,2}$}%
\affiliation{$^1$ AG Theoretische Quantenphysik, Theoretische Physik, Universit\"at des Saarlandes, D-66041 Saarbr\"ucken, Germany\\
$^2$ Departament de Fisica, Universitat Autonoma de Barcelona, E-08193 Bellaterra, Spain
}%


\date{\today}

\begin{abstract}
A cooling scheme for trapped atoms is proposed, which combines cavity-enhanced scattering and electromagnetically induced transparency. The cooling dynamics exploits a three-photon resonance, which combines laser and cavity excitations. It is shown that relatively fast ground-state cooling can be achieved in the Lamb-Dicke regime and for large cooperativity. Efficient ground-state cooling is found for parameters of ongoing experiments. 
\end{abstract}

\maketitle


\section{Introduction}

Control of the quantum dynamics of physical objects is a prerequisite for quantum technological applications. One important requirement is high-fidelity quantum state preparation. This often relies on cooling the physical system of interest to sufficiently low temperatures, such that a single quantum state, the ground state, is occupied with probability approaching unity. Laser cooling constitutes in this respect a successful technique, which is routinely used in the preparation stage of experiments with atoms and ions~\cite{Laser_Cooling,qo:eschner2003}. Furthermore, the various laser-cooling concepts and schemes which have been proposed and partly tested over the years are being considered for cooling molecules~\cite{Cool_Molecule2} and more complex objects, such as for instance phononic modes of micromechanical resonators to ultralow temperatures \cite{Aspelmeyer_JOSAB} and condensed-phase systems \cite{Cooling_Solids1,Cooling_Solids2}. 

The idea at the basis of laser cooling of trapped particles is to enhance scattering processes leading to a net transfer of the oscillator mechanical energy to the electromagnetic-field modes. Ground-state cooling of a harmonically-trapped particle, in particular, relies on a strong enhancement of the scattering processes which cool the oscillator, over the ones which heat the motion. Several ground-state cooling techniques have been discussed that apply this basic concept in various ways \cite{qo:eschner2003,Leibfried2003,Retzker1,Retzker2}. One scheme that is relevant to the study performed in this work is known in the literature as EIT-cooling, where EIT stands for Electromagnetically Induced Transparency \cite{EIT_Harris}, and it uses coherent population trapping \cite{arimondo:1996,EIT_Harris} due to quantum interference between laser-driven electronic transitions, in order to tailor the scattering cross section of the atoms to pursue ground-state cooling \cite{art:morigi2000,art:morigi2003}. The EIT-cooling scheme extends the basic concepts of velocity-selective coherent population trapping for free atoms \cite{VSCPTexp,VSCPT} to trapped atoms, and has been experimentally demonstrated in Refs.~\cite{EITexpPRL,EITexpAPB,kampschulte2010}. Further studies can be also found in Refs. \cite{Steane2004,evers:2004,Helm2008}.

A further tool that is being increasingly considered in order to enhance scattering processes leading to trapping and cooling of atoms, is the strong coupling of one or more atomic transitions to a high-finesse optical resonator  \cite{Horak97,Cool_Molecule1,Domokos03,Pinkse2000,Hood2000,VuleticCavCooling,vuletic2001,nussmann2005,murr2006}. For such a system it has been predicted that quantum interference effects can emerge from the quantized nature of the cavity mode \cite{Field1993,Rice1996}. When in addition the field is interfaced with the atomic vibration in a trap via the mechanical action of light, further interference effects can emerge that can increase the cooling efficiency \cite{Cirac95,Zippilli_PRL05}. The most recent observation of cavity-induced EIT~\cite{Rempe_CEIT,magnus2011}, and of mechanical effects associated with it \cite{Meschede,tan2011}, leads to the natural question whether and how EIT and cavity quantum electrodynamics can concur together to provide novel tools of control on the mechanical effects of atom-photon interactions. 

In this paper we present a theoretical study of the mechanical effects of light on a trapped atom in a setup which supports cavity-induced EIT. We show that the combination of EIT and cavity quantum electrodynamics can give rise to a cooling mechanism, which for an accurate choice of parameters allows one to prepare the atoms in the ground state of the potential with probability approaching unity. Remarkably, high efficiencies are found for the parameters of the experimental setup reported in Ref. \cite{Meschede}. We show that the cooling dynamics can be often explained by means of a three-photon resonance \cite{lukin:1999,Scully:3phot:1,Scully:3phot:2,champenois:2006}, which involves cavity and laser photons. For certain parameter choices cooling results from interference in the mechanical effects of the atomic interaction with the cavity and laser fields. 

This article is organized as follows. In Sec~\ref{Sec:Theory} the physical system and the theoretical model are introduced. The assumptions are discussed, which are at the basis of the theoretical treatment in this article. In Sec.~\ref{Sec:Dressed} the dark resonances and dressed states of the Hamiltonian for the electronic and cavity levels are reported, and the cavity and atom excitation spectra are discussed. In Sec. \ref{Sec:Cooling} the basic equations of cooling are derived, and in Sec. \ref{Sec:Results} the corresponding predictions are reported for  experimental parameters based on Ref.~\cite{Meschede}. The conclusions are drawn in Sec.~\ref{Sec:Conclusions}, while in the appendices the details of the calculations in Secs.~\ref{Sec:Theory}, \ref{Sec:Dressed}, and \ref{Sec:Cooling} are presented.

\section{The theoretical model}
\label{Sec:Theory}

\begin{figure}
\includegraphics[width=5cm]{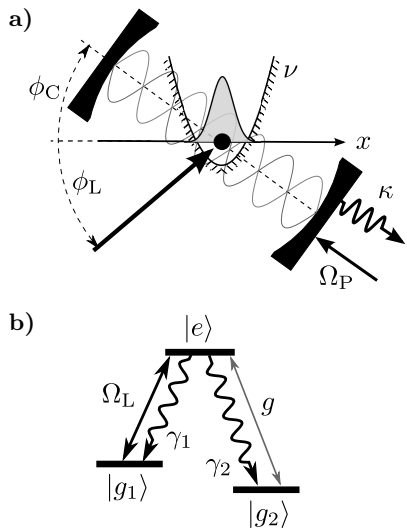}
\put(-150,190){\bf a)}
\put(-150,75){\bf b)}
\caption{\label{fig:setup} {\bf a)} Setup of the system. An atom is confined inside an optical resonator by a harmonic trap with frequency $\nu$. The atom is transversally driven by a laser at Rabi frequency $\Omega_{\rm L}$ and couples to the cavity mode with vacuum Rabi frequency $g$. The cavity mode is pumped by a laser (coupling strength $\Omega_\R{P}$) and decays with rate $\kappa$. $\phi_\R{L}$ ($\phi_\R{C}$) denotes the angle between the axis of the motion and the laser (cavity) wave vectors. {\bf b)} Relevant electronic transitions. The transverse laser (cavity mode) drives the transition $\ket{g_1} \to\ket e$ ($\ket{g_2} \to\ket e$). The excited state $\ket e$ decays spontaneously into the stable states $\ket{g_1}$ and $\ket{g_2}$  with rates $\gamma_1$ and $\gamma_2$, respectively.}
\end{figure}

A single atom of mass $M$ is confined inside of an optical resonator by  an external harmonic trapping potential and is irradiated by a laser field, while the cavity is pumped by a second laser field at strength $\Omega_\R{P}$. Two atomic dipolar transitions couple to the laser and the cavity mode, respectively, and share the same excited state, forming a $\Lambda$-shaped configuration of levels.  The atomic center-of-mass motion is treated in one dimension along the $x$ axis. Although the one-dimensional treatment seems to be a strong restriction, in the Lamb-Dicke regime which we will assume here, the rate equations for the cooling dynamics can be split into three independent sets of rate equation, one for each direction of motion. Thus, in this limit each spatial dimension can be treated separately~\cite{Eschner:2003}. Figure~\ref{fig:setup} illustrates the setup and the level configuration, highlighting the geometry of the laser beam and the cavity axis with respect to the axis of the motion. 

In the following we report some of the relevant parameters and introduce the basic notation. The atomic level configuration is reported in Fig.~\ref{fig:setup}(b). The atomic levels are denoted by the stable states $|g_1\rangle$ and $|g_2\rangle$ which are coupled by a dipolar transition with moments $\vec \wp_1$ and $\vec \wp_2$, respectively, to the common excited state $|e\rangle$. We denote the level frequencies by $\omega_1$, $\omega_2$ and $\omega_e$. The atomic dipole transition $\ket {g_1}\leftrightarrow\ket e$ is driven by a laser with frequency $\omega_{\rm L}$ and Rabi frequency $\Omega_\R{L}$. The transition  $\ket{g_2}\leftrightarrow\ket e$ is coupled to a mode of the optical resonator with frequency $\omega_\R{C}$, linewidth  $2\kappa$ and vacuum Rabi frequency $g$. The excited state has radiative linewidth $\gamma$ and decays into the state $\ket{g_j}$ with rate $\gamma_j$ ($j=1,2$) such that $\gamma=\gamma_1+\gamma_2$~\footnote{The assumption of a closed configuration is not necessary if an appropriate pumping scheme can be employed}. 

The atomic center-of-mass motion is confined by a harmonic potential of frequency $\nu$ which is independent on the electronic state of the atoms. This situation can be realized when the single particle is an ion in a Paul or Penning trap~\cite{IonTrap_Ghosh}, or a neutral atom when the confining potential is a dipole trap under specific conditions (such as the magic wavelength)~\cite{magic_wlNature,magic_wlScience}. The geometry of the setup is fixed by the angles $\phi_\R{L}$ and $\phi_\R{C}$ which give the orientations of the laser and cavity wave vectors with respect to the axis of motion, see Fig.~\ref{fig:setup}.

The dynamics of the system include the mechanical coupling of the atom with the electronic transition via absorption and/or emission of photons. Moreover, it describes the coupling of the cavity mode with the longitudinal modes of the electromagnetic field by the finite transmittance of the mirrors. Correspondingly, the Hamiltonian ${\mathcal H}$ governs the dynamics in the Hilbert space of the degrees of freedom of the atom, the cavity mode, and external modes of the electromagnetic field. It can be decomposed into the sum
\begin{equation}
{\mathcal H}=H+W_\gamma+W_\kappa+H_{\rm emf}\,.
\end{equation}
Here, $H$ is the Hamiltonian for the dynamics of the system, composed by cavity mode, electronic bound states and center-of-mass motion of the atom, and contains the coupling with the lasers, which are treated as classical fields. Term $H_{\rm emf}$ describes the Hamiltonian of the modes of the electromagnetic field external to the resonator. The coupling between the atomic dipole transitions and these modes is given by $W_\gamma$, which is responsible for the radiative instability of excited state $\ket{e}$ at rate $\gamma$. The longitudinal modes of the external electromagnetic field couple to the cavity mode via the finite mirror transmittance. This coupling is incorporated by the term $W_\kappa$  and gives rise to cavity losses at rate $\kappa$. In what follows we discuss in detail the form of the individual terms.

\subsection{Hamiltonian of the atom-cavity system}

The Hamiltonian
\begin{equation}
H = H_{\rm ext} + H_{\rm int} + H_{\rm cav} + W
\label{eq:H}
\end{equation}
governs the dynamics of the composite system of atom and cavity. In a frame rotating with the lasers' frequencies, the individual terms on the right-hand-side of Eq.~(\ref{eq:H}) read
\begin{align}
H_{\rm ext} &= \hbar\nu \left(b^\dagger b + \frac{1}2\right),\label{eq:Hext}\\
H_{\rm int} &= -\hbar\deltac\prj{e}+\hbar[\delta_1-\deltac]\prj{g_1}+\hbar\Delta\prj{g_2},\label{eq:Hint}\\
H_{\rm cav} &= -\hbar\Delta a^\dagger a\label{eq:Hcav},
\end{align}
and describe the center-of-mass motion of the atom in the harmonic potential of frequency $\nu$, the unperturbed dynamics of the internal electronic states, and the dynamics of the cavity mode, respectively. The operators $b$ and $b^\dagger$ in Eq.~\eqref{eq:Hext} annihilate or create single vibrational excitations and are connected with the atom's position operator by the relation
\begin{equation}
x=\xi(b+b^\dagger),
\end{equation}
where $\xi=\sqrt{\hbar/2 M \nu}$ denotes the size of the ground-state wave packet. The operators $a$ and $a^\dagger$ in Eq.~\eqref{eq:Hcav} annihilate and create, respectively, a photon in the considered mode, and
\begin{align}
\Delta=\omega_\R{P}-\omega_\R{C}\label{eq:Delta}
\end{align}
is the detuning between the cavity and the probe. The other detunings occurring in Eq.~\eqref{eq:Hint} are
\begin{align}
\delta_1 &= \omega_{\rm L} - (\omega_e - \omega_1)\label{eq:delta1}\,,\\
\deltac &= \omega_{\rm C} -(\omega_e - \omega_2).\label{eq:deltac}
\end{align}
The interaction part
\begin{align}
W = W_\R{P} + W_\R{L} + W_\R{C}
\end{align}
is composed of
\begin{align}
W_\R{P} &= \frac{\hbar\Omega_\R{P}}2\left[a + a^\dagger\right],\label{eq:WP}\\
W_\R{L}(x) &= \frac{\hbar\Omega_\R{L}}2
\left[
\ketbra{e}{g_1}e^{i k x \cos\phi_\R{L} }
+ {\rm H.c.}
\right],\label{eq:WL}\\
W_\R{C}(x) &= \hbar g(x)\left[\ketbra{e}{g_2} a + {\rm H.c.}\right]\label{eq:WC},
\end{align}
where $W_\R{P}$ is the drive of the cavity field by the probe of power $P$ with strength $\Omega_\R{P}=2\sqrt{P\kappa/\hbar\omega_\R{P}}$, $W_\R{L}(x)$ is the coupling of the atomic dipole $\ket{g_1}\leftrightarrow\ket e$ to the control laser with Rabi frequency $\Omega_\R{L}$, and $W_\R{C}(x)$ the Jaynes-Cummings interaction between the cavity mode and the transition $\ket{g_2}\leftrightarrow\ket e$ with coupling constant
\[
g(x)=g\cos(k x\cos\phi_{\rm C}+\varphi)\,.
\]
The quantity $k$ is the wavenumber of the control laser field and of the cavity mode \footnote{We assume that the two lower states ${\ket{g_j}}$ are atomic hyperfine states. Then, the wave numbers of cavity $k_{\rm C}$ and the lasers $k_{\rm L}$, $k_{\rm P}$ can be assumed to be approximately equal: $k\approx k_{\rm C}\approx k_{\rm L}\approx k_{\rm P}$. 
}. Moreover, $\varphi$ determines the equilibrium position of the atom in the trapping potential with respect to the cavity mode function.

\subsection{Coupling to the electromagnetic field external to the resonator}

We include the electromagnetic field outside the cavity into the theoretical model using a Hamiltonian description instead of a master equation. This provides more insight into the basic processes underlying the cooling dynamics. The Hamiltonian
\begin{subequations}
\label{eq:Hemfs}
\begin{align}
H_{\rm emf} = {\sum_{\vec k,\epsilon}}^{(\gamma_1)} \hbar[\omega_{\vec k}-\omega_\R{L}] c_{\vec k, \epsilon}^\dagger c_{\vec k,\epsilon}\nonumber\\
+{\sum_{\vec k, \epsilon}}^{(\gamma_2)} \hbar[\omega_{\vec k}-\omega_\R{P}] c_{\vec k, \epsilon}^\dagger c_{\vec k, \epsilon}\nonumber\\
+{\sum_{\vec k, \epsilon}}^{(\kappa)} \hbar[\omega_{\vec k}-\omega_\R{P}] c_{\vec k, \epsilon}^\dagger c_{\vec k, \epsilon}
\end{align}
\end{subequations}
accounts for the energy of the transversal and longitudinal field modes which couple independently to the atomic dipoles and the cavity, respectively.  The superscript $\gamma_j$ ($\kappa$) indicate that the sum is restricted to the modes which couple quasi-resonantly with the transition $|g_j\rangle\to |e\rangle$ (with the cavity mode) and the Hamiltonian is reported in the reference frame at which the corresponding transition rotates at the frequency of the driving laser. The quantity $\omega_{\vec k}=c|\vec k|$, with the speed of light in vacuum $c$, denotes the frequency of the electromagnetic field mode, and $c_{\vec k,\epsilon}$ is the annihilation operator of the mode labeled by the wavevector $\vec k$ and polarization $\vec e_{\vec k,\epsilon} \perp \vec k$ with $\epsilon = 1,2$.

The interaction between the atomic dipole $\ket{g_j}\leftrightarrow \ket{e}$ and the external electromagnetic field reads
\begin{align}
W_{\gamma}(x) &=W_{{\gamma_1}}(x) + W_{{\gamma_2}}(x)\nonumber\\
&=\sum_{j=1,2}{\sum_{\vec k,\epsilon}}^{(\gamma_j)}\hbar
\left[ g_{\vec k,\epsilon}^{(\gamma_j)} \ketbra{e}{g_j}e^{i(\vec k\cdot\vec e_x) x}c_{\vec k,\epsilon}+{\rm H.c}\right]
\label{eq:Wgammaj}
\end{align}
where the coupling constant $g_{\vec k,\epsilon}^{(\gamma_j)}=\vec\wp_j \cdot\vec e_{\vec k,\epsilon}{\mathcal E}_{\vec k}/\hbar$ is proportional to the scalar product of the atomic dipole moment $\vec\wp_j$ with the vacuum electric-field ${\mathcal E}_{\vec k}\vec e_{\vec k,\epsilon}$. The coupling of the cavity mode to the longitudinal modes of the external electromagnetic field is given by
\begin{align}
W_\kappa &= \hbar {\sum_{\vec k,\epsilon}}^{(\kappa)}
\left[g_{\vec k,\epsilon}^{(\kappa)}  a^\dagger c_{\vec k,\epsilon}+ {\rm H.c}\right]\,,
\label{eq:Wkappa}
\end{align}
with coupling constant $g_{\vec k,\epsilon}$ (a detailed form in terms of the physical parameters can be found for instance in Ref.~\cite{Carmichael}). We remark that the operator $W_\kappa$ does not depend on the atomic motion, but describes dissipation processes which will be instrumental for cooling the atomic motion~\cite{Domokos03}. We furthermore emphasize that quantum noise due to fluctuations of the atomic dipoles and the cavity field is systematically incorporated in the theoretical description by the couplings Eqs.~\eqref{eq:Wgammaj} and \eqref{eq:Wkappa}.

\subsection{Basic assumptions and perturbative expansion}

Throughout the paper we focus on the regime where: {\it (i)} The atom is tightly confined: The size of the atomic wavepacket $\Delta x$ is much smaller than the lasers' wavelengths. This is the so-called Lamb-Dicke regime \cite{qo:stenholm1986,qo:eschner2003}. It is found when the inequality $k\Delta x\ll 1$ is satisfied. Using that $\Delta x= \xi\sqrt{2\langle m\rangle+1}$ by averaging over a thermal state with mean occupation number $\langle m\rangle$, the inequality can be rewritten as
\begin{equation}
\eta\sqrt{2\langle m\rangle +1}\ll 1\label{eq:condLD}
\end{equation}
where $\eta=k\xi$ is the Lamb-Dicke parameter, which scales the mechanical effects of light on the atomic motion~\cite{qo:stenholm1986}. For later convenience we define
\begin{align*}
\eta_\R{L} &= \eta \cos\phi_\R{L}\,,\\
\eta_\R{C} &= \eta \cos\phi_\R{C}\,,
\end{align*}
which account for the geometry of the mechanical coupling of laser and cavity to the axis of motion, respectively. 

We further assume that {\it (ii)} the laser driving the cavity is sufficiently weak such that the average photon number of the cavity mode is much smaller than unity. This corresponds to taking the small parameter
\begin{equation}
|\epsilon|^2 \equiv \left|\frac{\Omega_{\rm P}/2}{\Delta+i\kappa}\right|^2\ll 1\,,\label{eq:condWP}
\end{equation}
which gives the mean number of intracavity photons when no atom is present. The requirement \eqref{eq:condWP} can be achieved for high-finesse cavities by adjusting the parameters $\Omega_{\rm P}$ and $\Delta$ accordingly.

This regime allows for a perturbative treatment of the dynamics. We expand the total Hamiltonian into different orders of the Lamb-Dicke parameter by performing a Taylor expansion in $\eta$ of the exponentials $\exp[\dots x]$ in Eqs.~\eqref{eq:WL} and \eqref{eq:Wgammaj} and the function $g(x)$ in Eq.~\eqref{eq:WC}~\cite{Vitali06}. Moreover, we separate the weak driving term $W_{\rm P}$, {\it i.e.}, the first order term in $|\epsilon|$, from ${\mathcal H}_0$, so that the Hamiltonian takes on the form
\begin{align}
{\mathcal H} = {\mathcal H}_0 + W_{\rm P} + {\mathcal H}_{1} + {\rm O}(\eta^2)\,.
\end{align}

The term $ {\mathcal H}_0 $ is in lowest order in $\eta$ and $|\epsilon|$, and reads
\begin{align}
{\mathcal H}_0 = H_0 + W_{\gamma}(0) + W_{\kappa}\,.
\end{align}
Here,
\begin{align}
H_0 = H_{\rm ext} + H_{\rm opt} + H_{\rm emf}\,,
\label{eq:H0}
\end{align}
and
\begin{align}
H_{\rm opt} = H_{\rm int} + H_{\rm cav} + W_{\rm L}(0) + W_{\rm C}(0) \,,
\label{eq:Hopt}
\end{align}
which is given in zero order in the perturbative expansion. In this order of perturbation theory, the stable state of the atom-cavity system is
\begin{align}
\ket{\Psi_{\rm st}} = \ket{g_2,0}\,,
\label{eq:ketg20}
\end{align} 
which is the product state of the atom in $\ket{g_2}$ and no cavity photon, $|n=0\rangle$. Under the influence of $H_0$ (assuming the regime in which the Wigner-Weisskopf approximation can be applied \cite{Carmichael}), the atom-cavity system relaxes into the state $\ket{\Psi_{\rm st}}$, while the atomic center-of-mass motion evolves coherently and is decoupled from the electronic dynamics.

Coupling between the center-of-mass motion and the light is introduced in first order in the Lamb-Dicke parameter $\eta$. The corresponding term of the total Hamiltonian reads
\begin{align}
{\mathcal H}_1 = F x + \sum_{j=1,2} F_{\gamma_j} x\,.
\end{align}
It accounts for the mechanical effects due to absorption and emission of photons from/to the control laser and the cavity mode, where the operator $F$ 
\begin{align}
F=&F_{\rm L} + F_{\rm C}\nonumber\\
=&\frac{d}{d x}\left.\left(W_\R{L}(x) + W_\R{C}(x)\right) \right|_{x=0},
\label{eq:force}
\end{align}
can be interpreted as a force operator~\cite{Nienhuis}, while
\begin{align}
F_{\gamma_j} = \left.\vec e_x \cdot \left[\vec\nabla W_{\gamma_j}(x)\right]\right|_{x=0}
\label{eq:Fgammaj}
\end{align}
gives rise to the stochastic force, which is associated with the recoil due to spontaneous emission~\cite{Nienhuis}. Here, $\vec e_x$ is the unit vector along the axis of the motion.

The Lamb-Dicke regime is found when the condition set by Eq.~\eqref{eq:condLD} is fulfilled, and it allows one to assume the separation of time scales which determine the center-of-mass motion of the atom, and the common, internal-state dynamics of the atom-cavity system. At lowest order in the Lamb-Dicke expansion, the internal and external atomic degree of freedom evolve independently. The relevant time scale is given by the smallest  relaxation time of the atom-cavity system. At higher order, the atomic center-of-mass motion experiences the force due to the gradient of the electromagnetic field over the extension of its wave packet. These processes take place on a time scale, which is slower by a factor $\eta^2$ with respect to the time scale of the internal motion. In this regime the internal motion follows the external motion adiabatically~\cite{qo:stenholm1986,qo:cirac1992,art:morigi2003}. It is therefore instructive to first study the internal dynamics, neglecting the coupling with the external motion: This permits one to identify scattering processes which leads to cooling. 

\section{Level structure and properties of the atom-cavity system}
\label{Sec:Dressed}

\begin{figure}
\centerline{\includegraphics[width=7cm]{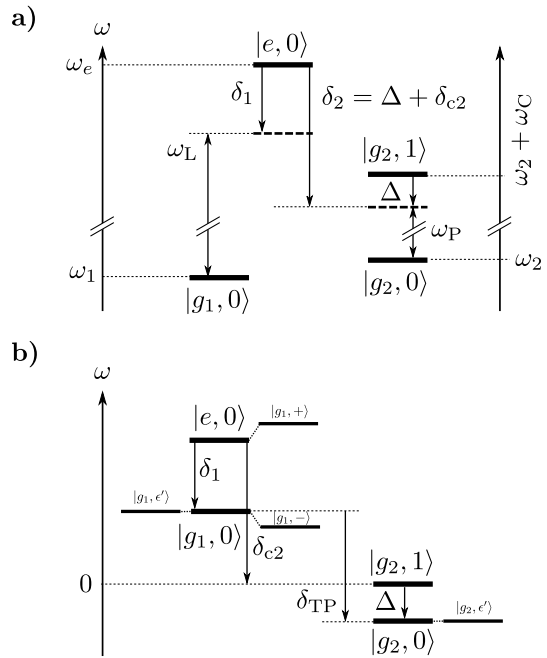}}
\caption{\label{fig:levelscheme} Level scheme of the atom-cavity system in a) laboratory and b) rotating frame. The vertical axis gives the frequency $\omega$. Downward arrows denote negative detuning. In b) the dressed states $\ket{g_1,\pm}$  and the states $\ket{g_2,\epsilon'}$ with their corresponding frequency shifts are reported (see text).  The detunings are defined in Eqs.~\eqref{eq:Delta}-\eqref{eq:deltac} and \eqref{eq:tpr}.}
\end{figure}

We now discuss the dressed states of Hamiltonian  $H_{\rm opt}$, Eq.~(\ref{eq:Hopt}). Condition~\eqref{eq:condWP} allows one to restrict the cavity-atom Hilbert space to states which contain at most one excitation of the cavity mode. The dynamics then takes place within the subspace spanned by the states  
\[
\{\ket{g_2,1}, \ket{e,0}, \ket{g_1,0}, \ket{g_2,0}\}\,. 
\]
These coupled levels constitute an effective four-level system, which is depicted in Fig.~\ref{fig:levelscheme}.  Such configuration of levels has been studied in the literature for the case in which the states are four different electronic levels, that are coupled by classical laser fields~\cite{lukin:1999,Scully:3phot:1,Scully:3phot:2,yelin:2003,champenois:2006}. For this case it has been found that this level scheme can exhibit dark resonances due to quantum interference between excitation paths involving three photons. The purpose of this section is to analyze the spectroscopic properties of the four-level system arising from the atom-cavity coupling, in the regime in which we can neglect the coupling with the center-of-mass motion. We will focus, in particular, on the conditions under which dark states exist as they will turn out to be relevant for the cooling dynamics, and refer the reader to Ref. \cite{champenois:2006} for an extensive analysis. 

\begin{figure}
\centerline{\includegraphics[width=7.5cm]{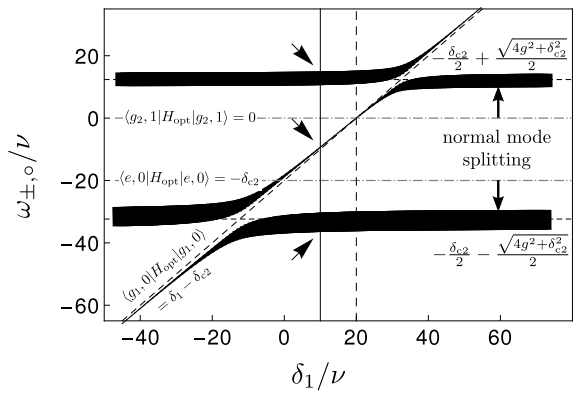}}
\caption{\label{fig:dstates} Frequencies and linewidths of the dressed states of the atom-cavity system as a function of $\delta_1$. The frequencies are found diagonalizing 
$H_{\rm opt}$, Eq.~(\ref{eq:Hopt}), over the basis of the unperturbed states $\{\ket{e,0}, \ket{g_1,0}, \ket{g_2,1}\}$, and correspond to the centers of the black curves. The corresponding linewidths determine the curves' width.  The horizontal dashed-dotted lines and the dashed diagonal line give the frequencies of the unperturbed states as a function of $\delta_1$. The arrows mark the frequencies of the resonances shown in Fig.~\ref{fig:excspec}. The parameters are $\kappa = 2\nu$, $\gamma = 10\nu$, $g=20\nu$, $\Omega_{\rm L} = 12\nu$, $\varphi = 0$, $\delta_{{\rm c}2} = 20\nu$.}
\end{figure}

The eigenspectrum of $H_{\rm opt}$, Eq.~(\ref{eq:Hopt}), within the considered subspace is displayed in Fig. \ref{fig:dstates} as a function of the frequency of the transverse laser, under the assumption that the state $\ket{g_2,0}$ is weakly coupled to the other states. The frequencies of the states $\ket{e,0}, \ket{g_1,0}, \ket{g_2,1}$, which diagonalize $H_{\rm opt}$ when the coupling to laser and cavity is set to zero, are indicated by the dashed curves. In presence of the coupling with the fields, the frequencies are shifted and degeneracies become avoided crossings. The resulting eigenfrequencies are $\omega_\pm$, $\omega_\circ$, which correspond to the eigenvectors $\ket \pm$ and $\ket \circ$, being superposition of the three states $\ket{e,0}, \ket{g_1,0}, \ket{g_2,1}$. The horizontal part of the curves, in particular, denote the frequency of the dressed states of the Jaynes-Cummings Hamiltonian, namely superposition of states $\ket{g_2,1}$ and $\ket{e,0}$. These states appreciably mix with state $\ket{g_1,0}$ at the level crossing with the curve $\delta_1-\deltac$. This is also visible by inspecting the linewidth, indicated in the figure by the breadth of the curves.

We now analyze the excitation spectra of cavity and atom, namely, the rate of photon emission by the cavity and the atom, as a function of the probe frequency $\Delta$. They are evaluated for the stationary state of the (internal) atom-cavity system, and are proportional to the probability that the cavity contains one photon and that the atom is in the excited state, respectively. In Appendix A a detailed calculation is reported, which shows that they take on the form
\begin{subequations}
\begin{align}
S_{\rm exc}^\kappa(\Delta) &\propto |\cF_\kappa(\Delta)|^2\label{eq:Sexcgamma}\,,\\
S_{\rm exc}^\gamma(\Delta) &\propto \ |\cF_\gamma(\Delta)|^2\label{eq:Sexckappa}
\end{align}
\label{eq:excspecs}
\end{subequations}
with
\begin{subequations}
\begin{align}
\cF_\kappa(\Delta) &= \frac{(\deltac+\Delta-\delta_1)\left[\deltac+\Delta+i\frac{\gamma_2}{2}\right]-\frac{\Omega_{\rm L}^2}{4}}{f(\Delta)},
\label{eq:cFkappa}
\\
\cF_\gamma(\Delta) &= g\frac{\deltac+\Delta-\delta_1}{f(\Delta)},
\label{eq:cFgamma}
\end{align}
\label{eq:cF}
\end{subequations}
where the superscript $\kappa$ ($\gamma$) indicates that it refers to cavity (atom) emission, see App.~\ref{app:scatter} for details. The relevant resonances can be identified with the poles of the 
function $f(\Delta)$, that correspond to the frequencies of the dressed states of $H_{\rm opt}$. 

\begin{figure}
\centerline{\includegraphics[width=7cm]{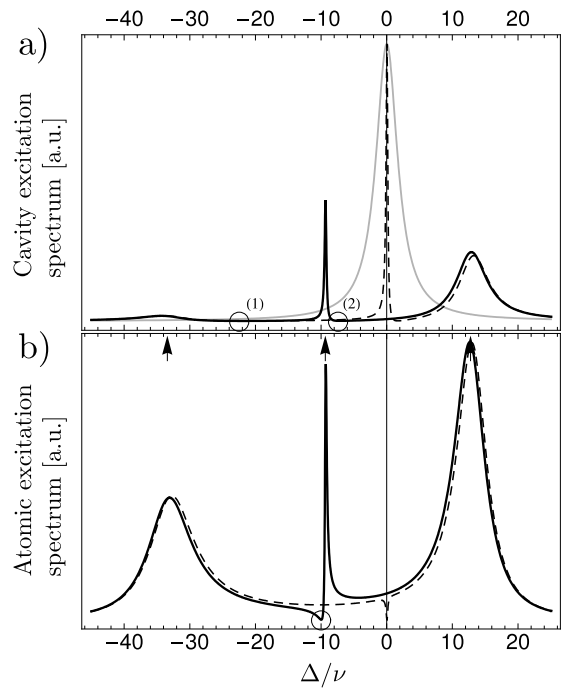}}
\caption{\label{fig:excspec} Excitation spectra of {\bf a)} cavity and {\bf b)} atom, in arbitrary units, as a function of the probe frequency $\Delta$, in units of $\nu$, for $\deltac=20\nu$ and $\delta_1=10\nu$ (black solid line) and $\delta_1=\deltac=20\nu$ (dashed line). The gray curve in {\bf a)} gives the Lorentzian excitation spectrum of the unperturbed cavity and is plotted for comparison. The spectra consist of three peaks at the frequencies of the dressed states of the manifold $\{\ket{e,0}, \ket{g_1,0},\ket{g_2,1}\}$, which are marked with arrows. The circles mark the frequencies for which the excitation spectra exhibits local minima, corresponding to approximate dark resonances.  In {\bf b)} the atomic excitation spectrum vanishes at $\Delta = \delta_1-\deltac$ (small circle), due to three-photon resonance between $\ket{g_1,0}$ and $\ket{g_2,0}$. The other parameters are $\kappa=2\nu$, $\gamma=10\nu$, $g=20\nu$, $\Omega_{\rm L}=12\nu$, $\varphi=0$.}
\end{figure}

The excitation spectra for cavity and atom are displayed in Fig.~\ref{fig:excspec}a) and b), respectively, as a function of $\Delta$. We recall that in absence of the atom, the cavity excitation spectrum is a Lorentz curve centered at the mode frequency with full width $2\kappa$, and corresponds to the grey curve reported in the figure. The solid curve corresponds to the cavity excitation spectrum in presence of the atom. The three peaks are identified with the dressed states of the system. Their centers are marked by the arrows in the abscissa and correspondingly by the arrows in Fig.~\ref{fig:dstates}. These peaks are also present in the atomic excitation spectrum, Fig.~\ref{fig:excspec}b). Additionally, we observe 
that the spectrum vanishes at a value of the probe frequency, and in this frequency region it exhibits a Fano-like profile. This is verified when the three-photon resonance condition is fulfilled, namely, when states $\ket{g_2,0}$ and  $\ket{g_1,0}$ are resonantly coupled, and corresponds to 
\begin{align}
\delta_{\rm TP} \equiv \deltac + \Delta - \delta_1 =0.
\label{eq:tpr}
\end{align}
Moreover, the cavity excitation spectrum exhibits two points, where the cavity response is minimal. These minima are marked with two circles in Fig. \ref{fig:excspec}a). 

The Fano-like profile and the minima in the cavity excitation spectrum can be understood in terms of quantum interference between the dressed states and state $\ket{g_2,0}$, which can give rise to significant effects even though state $\ket{g_2,0}$ is very weakly coupled to the other levels. In order to gain more insight, we first analyze  the atom cavity levels in a unitarily equivalent scheme, where $H_{\rm cav}+W_\R{P}$ is diagonal. We consider the states
\begin{align}
 \ket{g_j,\epsilon'} = D_{\rm c}(\epsilon')\ket{g_j,0}\quad (j=1,2)
 \label{eq:dispstates}
\end{align}
with the displacement operator $D_{\rm c}(\epsilon) = \exp(\epsilon a^\dagger - \epsilon^\ast a - i\Delta|\epsilon|^2 t)$ for the cavity mode and $\epsilon'=\Omega_{\rm P}/2\Delta$. For $\deltac + \Delta - \delta_1 = 0$, the states Eq.~\eqref{eq:dispstates} are resonantly coupled, such that 
\begin{align}
\ket{D} \propto \epsilon' g\cos\varphi\ket{g_1,\epsilon'}-\frac{\Omega_{\rm L}}{2} \ket{g_2,\epsilon'}
\end{align}
is an eigenstate of the Hamiltonian $H_{\rm opt}$, that has zero projection onto the electronic excited state. In the limit in which the pump is weak and out of resonance the state $|D\rangle$ is to good approximation the stationary state of the system and is a dark state. Equation~\eqref{eq:tpr} then gives the correct resonance condition and explains the position of the minimum in the atomic excitation spectrum.
Similarly, the cavity excitation spectrum exhibits minima when the state $\ket{g_2,0}$ is resonantly coupled to one of the dressed states $\ket{g_1,\pm}$, which diagonalize the laser interaction $H_{\rm int}+W_{\rm L}(0)$. 

Another dark resonance is found when the states  $\ket{g_2,1}$ and $\ket{g_1,0}$ are resonantly coupled, namely, when $\delta_1 = \deltac$. We denote this situation by ``two-photon resonance''. 
In this case the state 
\begin{align}
\ket{D_{\cM}}\propto g\cos\varphi\ket{g_1,0}-\frac{\Omega_{\rm L}}{2}\ket{g_2,1}
\end{align}
is an eigenstate of the Hamiltonian $H_{\rm opt}$ and is stable over a time scale, in which cavity decay can be neglected. This situation is reported by the dashed curves in Figs. ~\ref{fig:excspec} a) and b). In this case, the atomic excitation spectrum exhibits only two peaks, and the dark resonance is at the frequency where the cavity output is maximal.

In the next section we take into account the motion of the atom, and trace back the properties of the rates of heating and cooling transitions to the characteristics identified in the excitation spectra.

\section{Theory of cooling in the Lamb-Dicke limit}
\label{Sec:Cooling}

In the Lamb-Dicke regime, the atom's external and internal degrees of freedom are weakly coupled. In this limit, one can assume that the system, composed by cavity and electronic excitations of the atom, reaches a steady state over a time scale which is much faster than the one at which the motion evolves. The corresponding theoretical treatment has been discussed in extent, for instance in Ref. \cite{qo:cirac1992,art:morigi2003,Zippilli_PRA05}. This justifies the formulation of the dynamics of the external degrees of freedom in the form of a rate equation for the occupations $p_m(t)$ of the vibrational state $\ket{m}$, which reads
\begin{align}
\frac{d p_m}{d t} = &(m+1) A_- p_{m+1} - [(m+1)A_+ + m A_-] p_m \nonumber\\ &+ m A_+ p_{m-1}\,.
\label{eq:rateeqcooling}
\end{align}
The average phonon number obeys the equation $\langle m\rangle(t) = -\Gamma\langle m\rangle(t) + A_+$ with the cooling rate
\begin{equation}
\Gamma = A_- - A_+\,.
\label{eq:coolingrate}
\end{equation}
When $\Gamma>0$, {\it i.e.}, $A_->A_+$, a stationary state exists.  At steady state, the flow of population fulfills the detailed balance condition, and the mean occupation at steady state reads~\cite{qo:englert2002,qo:stenholm1986}
\begin{equation}
p_m^{\rm st} = \left(1-\frac{A_+}{A_-}\right) \left(\frac{A_+}{A_-}\right)^{m}
\end{equation}
with mean phonon occupation 
\begin{equation}
\langle m\rangle_{\rm st} = \frac{A_+}{A_--A_+}.
\label{eq:meanm}
\end{equation}
These formula show that high cooling rate are reached by maximizing the value of $A_-$ as well as the ratio $A_-/A_+$, while the ratio $A_-/A_+$ alone controls the final temperature. Various cooling schemes resort to different control tools in order to achieve either fast cooling and/or low temperatures. In the following we will characterize the dynamics which can lead to efficient ground-state cooling in this setup.

\subsection{Cooling and heating rates}

\begin{figure*}
\centerline{\includegraphics[width=15cm]{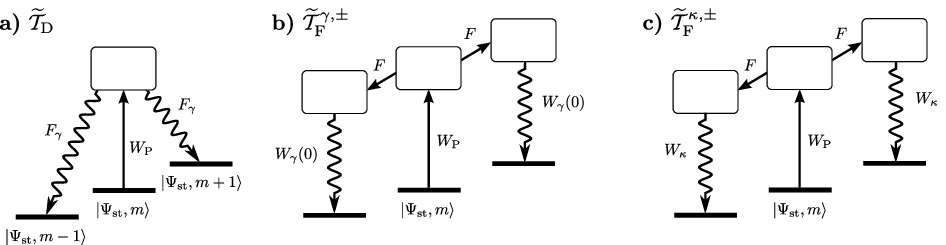}}
\caption{\label{fig:sp} Schematic representation of the scattering processes leading to a change of the motional state, corresponding to Eqs.~\eqref{eq:tTD}, \eqref{eq:tTL} and \eqref{eq:tTC}.  The arrows show the sequence of processes leading to a transition $\ket{\Psi_{\rm st},m}\to\ket{\Psi_{\rm st},m\pm 1}$. The intermediate internal states are denoted by the rounded box. Process in (a) terminates by a spontaneous decay, in which the mechanical effect is due to the spontaneously scattered photon. In (b) and (c) the transition is terminated by emission of a photon by spontaneous emission and cavity decay, respectively, and the  mechanical effects are due either by absorption or emission of a photon of the control laser ($F=\R{L}$) or of the cavity mode ($F=\R{C}$).}
\end{figure*}

The transition rates $A_\pm$ in Eq. (\ref{eq:rateeqcooling})  can be calculated in perturbation theory using the resolvent of the Hamiltonian. We refer the reader to App.~\ref{app:scatter} for the detailed evaluation. They describe absorption of a photon of the probe laser pumping the cavity, followed by emission into the modes of the external field either by atomic or by cavity decay, and can be written as the sum of various contributions of different physical origin:
\begin{eqnarray}
A_\pm &=& \diffD+ \sum_{r=1,2}\gamma_r|\widetilde \cT^{\gamma_r,\pm}_{\rm L}+\widetilde \cT^{\gamma_r,\pm}_{\rm C}|^2\nonumber\\
& &+2\kappa|\widetilde \cT^{\kappa,\pm}_{\rm L}+\widetilde \cT^{\kappa,\pm}_{\rm C}|^2\,.
\label{eq:ApmT}
\end{eqnarray}
The quantity $\diffD$ is a diffusion coefficient, originating from the mechanical effects of the spontaneously emitted photon, while $\widetilde \cT_{F}^{\ell,\pm}$ denote the transition amplitude associated with the absorption ($-$) or emission ($+$) of a vibrational phonon by either scattering the probe photon into the external modes by cavity decay ($\ell=\kappa$) or by spontaneous emission along one of the two transitions ($\ell=\gamma_{1,2}$). The subscript $F$ indicates whether the mechanical effect is due to the laser photon ($F=\R{L}$) or to the cavity photon ($F=\R{C}$). For a given scattering process, these mechanical effects can interfere constructively, as one observes from the fact that the corresponding contribution, $|\widetilde \cT^{\ell,\pm}_{\rm L}+\widetilde \cT^{\ell,\pm}_{\rm C}|^2$, is the coherent sum of the individual amplitudes. A schematic representation of the corresponding scattering processes is depicted in Fig. \ref{fig:sp}. 

A similar decomposition, as the one in Eq. \eqref{eq:ApmT}, was identified for the heating and cooling rates of a trapped atom, whose two-level, dipole transition couples simultaneously to the cavity mode and to a transversal laser \cite{Zippilli_PRA05}.  In such a setup several interference processes have been identified which can be tuned in order to enhance the cooling efficiency. In the present system, where an additional electronic transition is involved, a further interference process can take place, which is analogous to EIT.  

Below we discuss in detail the individual terms for the system considered in this work. In the following we neglect  the decay along transition $\ket{e}\to\ket{g_1}$, and take $\gamma_2=\gamma$, unless otherwise specified. This assumption leads to a considerable simplification of the analytical expressions we are going to report and does not affect qualitatively the cooling dynamics for the parameter regime here explored, as numerical checks show.  

We first consider the diffusion term $\diffD$ in Eq. \eqref{eq:ApmT}. This term describes scattering processes where a probe photon is spontaneously emitted by the atom, and where the mechanical effect is due to photon recoil by spontaneous decay. It can be written in the form 
\begin{equation}
\diffD=\gamma\Wgeo{2}|\widetilde \cT_{\rm D}|^2\,,
\end{equation}
where $\Wgeo{2}$ is a geometrical factor which depends on the atomic dipole pattern of radiation and
\begin{align}
\widetilde\cT_{\rm D} = -i\eta \frac{\Omega_{\rm P}}2 \cos\varphi \cF_\gamma(\Delta)
\label{eq:tTD}
\end{align}
is the transition matrix element (apart of some constant factors), where the function $\cF_\gamma(\Delta)$ is given in Eq.~\eqref{eq:cFgamma}. Hence, $\diffD$ is proportional to the atomic excitation spectrum in Eq. \eqref{eq:Sexcgamma}, and thus to the probability that the excited state is occupied. In particular, it vanishes at a node of the cavity mode function where $\cos\varphi=0$, since there, the state $\ket {g_2}$ does not couple to the excited state in zero order in the Lamb-Dicke expansion. Moreover, for general values of $\varphi$, $\diffD$ becomes zero when the atomic excitation spectrum vanishes, which is verified when $\Delta=\Delta_0$, with 
\[
\Delta_0\equiv \delta_1-\delta_{c2}\,, 
\]
namely, when states $\ket{g_2,0}$ and $\ket{g_1,0}$  are resonantly coupled. For these parameters the population of the atomic excited state is zero because of destructive interference between the excitation paths. Consequently, diffusion can be suppressed. This property is also at the basis of the so-called EIT-cooling mechanism \cite{art:morigi2003}. Differing from EIT cooling, suppression of diffusion is here due to a three-photon interference process.

We now turn our attention to the transition amplitudes 
\begin{subequations}
\label{eq:tTL}
\begin{align}
\widetilde\cT_{\rm L}^{\gamma,\pm} &= \mp i\eta_\R{L}\frac{\Omega_{\rm P}}{2}\frac{\Omega_{\rm L}^2}{4}\frac{ \nu (i\kappa+\Delta\mp\nu)g\cos\varphi}{f(\Delta\mp\nu)f(\Delta)}\,,
\label{eq:tTLg}
\\
\widetilde\cT_{\rm L}^{\kappa,\pm} &= \mp i \eta_\R{L}\frac{\Omega_{\rm P}}{2}\frac{\Omega_{\rm L}^2}{4}\frac{ \nu g^2\cos^2\varphi}{f(\Delta\mp\nu)f(\Delta)}\,.
\label{eq:tTLk}
\end{align}
\end{subequations}
They describe the scattering processes where absorption or emission of a photon of the transverse laser leads to a change in the vibrational motion and are depicted in Fig. \ref{fig:sp}b) and c).  These scattering amplitudes vanish at the node of the mode function where $\cos\varphi=0$, and when the control laser is perpendicular to the axis of motion ($\eta_\R{L}=0$), namely, when there is no mechanical effect of the control laser along the direction of the motion. When this is not verified, these amplitudes may become maximal when the parameters are chosen so that the energy defect of the intermediate scattering states becomes minimal. Cooling transitions are enhanced over heating transitions, in particular, for the parameters for which equation
\begin{equation}
\label{f:Delta}
{\rm Re}\{f(\Delta+\nu)\}=0
\end{equation}
is fulfilled.

Finally, the transition amplitudes 
\begin{subequations}
\label{eq:tTC}
\begin{align}
\widetilde\cT_{\rm C}^{\gamma,\pm} &= -\eta_{\rm C}\frac{\Omega_{\rm P}}{2} \sin\varphi\, \cF_\gamma(\Delta\mp\nu)\,\times\nonumber\\
&\big[g \cos^2\varphi \, \cF_\gamma(\Delta) + (i\kappa+\Delta\mp\nu)\, \cF_\kappa(\Delta)\big]\,,
\label{eq:tTCg}
\\
\widetilde\cT_{\rm C}^{\kappa,\pm} &= -\eta_{\rm C}\frac{\Omega_{\rm P}}{2} \frac{\sin2\varphi}2\times\nonumber\\ &g\big[\cF_\gamma(\Delta\mp\nu)\,\cF_\kappa(\Delta)+\cF_\gamma(\Delta)\,\cF_\kappa(\Delta\mp\nu)\big]
\label{eq:tTCk}
\end{align}
\end{subequations}
describe the scattering processes where absorption or emission of a photon of the cavity field leads to a change in the vibrational motion. They are depicted in Fig. \ref{fig:sp}b) and c).  
These transition amplitudes vanish when the cavity axis is perpendicular to the axis of motion ($\eta_{\rm C}=0$), namely, when there is no mechanical effect of  the cavity photon along the direction of the motion, or alternatively when the atom is situated at an antinode of the mode function, so that the derivative of the cavity-ion potential vanishes and the corresponding force operator is zero. Term $\widetilde\cT_{\rm C}^{\kappa,\pm}$, Eq.~\eqref{eq:tTCk}, also disappears at a node of the cavity standing wave. Interestingly, both terms are products of the Fano-like factors $\cF_\gamma$, Eq.~\eqref{eq:cFgamma}, connected with the atomic excitation spectrum, and $\cF_\kappa$, Eq.~\eqref{eq:cFkappa}, which reflects the characteristics of the excitation spectrum $S_{\rm exc}^\kappa$, Eq.~\eqref{eq:Sexckappa}.

We now identify some conditions, for which ground-state cooling can be implemented.

\subsection{Three-photon resonance: EIT cooling and beyond}
\label{sec:tpr}

In this section we study the cooling dynamics under the assumption of three-photon resonance, 
\[
 \delta_{\rm TP}=\Delta+\deltac-\delta_1 = 0\,,
\]
 corresponding to setting $\Delta=\Delta_0$. In this case the transition rates simplify considerably: Due to destructive interference of the excitation paths, the excited state $\ket e$ is not occupied and the diffusion term $\diffD$ in Eq. \eqref{eq:ApmT} vanishes. This property also leads to a considerable simplification of the transition amplitudes $\widetilde T_{\rm C}^{\ell,\pm}$, since $\cF_\gamma(\Delta_0)=0$. In this case the rates take on the form \footnote{We note that Eq.~\eqref{eq:ApmEIT} and the following formulas derived from it acquire the same form also for $\gamma_1\neq 0$: The parameter $\gamma$ appearing there is the total radiative linewidth of the excited state.} 
\begin{align}
A_\pm =& |\epsilon|^2\widetilde\eta^2 g^2\gamma\times\nonumber\\
&\frac{1+\mathcal C_\pm}
{\frac{\gamma^2}{4}\left(1+\mathcal C_\pm\right)^2 + \left(\frac{\Omega_\R{L}^2}{4\nu}-\nu\pm \delta_1+
\frac{\gamma}{2\kappa}\mathcal C_\pm(\nu \mp \Delta)\right)^2}\,,
\label{eq:ApmEIT}
\end{align}
where 
\begin{align}
\widetilde\eta^2 = \eta^2_{\rm L}\cos^2\varphi +  \eta^2_{\rm C}\sin^2\varphi
\end{align}
contains the dependency on the geometry of the setup, and the parameters 
\begin{equation}
\mathcal C_\pm=C \frac{\kappa^2}{\kappa^2+(\Delta\mp\nu)^2}
\end{equation} 
are proportional to the single-atom cooperativity 
\begin{align}
C=\frac{g^2\cos^2\varphi}{\kappa\gamma/2}.
\label{eq:singleC}
\end{align}
The rate of cooling (heating) is maximum, when state $\ket{g_2,0}$ is set on resonance with the red (blue) sideband of one of the dressed states of $H_{\rm opt}$. Cooling is enhanced, for instance, when the denominator of rate $A_-$ in Eq.~\eqref{eq:ApmEIT} becomes minimal, which is satisfied when relation
\begin{align}
\delta_1=\frac{\Omega_{\rm L}^2}{4\nu}-\nu+(\nu+\Delta)\frac{g^2\cos^2\varphi}{\kappa^2+(\Delta+\nu)^2}
\label{eq:rescondeit}
\end{align}
is fulfilled.

\subsubsection{Cavity EIT cooling}
 
In order to acquire a better understanding on the role of the resonator and identify the regimes in which cooling occurs,  we first analyze the case $\mathcal C_\pm\ll 1$. The parameters ${\mathcal C}_\pm$ become small for {\it (i)} small cooperativities $C$, {\it (ii)} when the pump on the cavity is far-off resonant, or {\it (iii)} for $\kappa\ll |\Delta\pm\nu|$.
Then, the rates are given by
\begin{align}
A_\pm^{\ll} =& |\epsilon|^2\widetilde\eta^2\times\nonumber\\
&\frac{\gamma g^2\nu^2}
{\nu^2\frac{\gamma^2}{4} + \left[\frac{\Omega_\R{L}^2}{4}-\nu(\nu\mp \delta_1)+
\frac{\gamma \kappa}{2} C \frac{\nu(\nu\mp \Delta)}{\kappa^2+(\Delta\mp \nu)^2}\right]^2}.
\label{eq:Apmasym}
\end{align}
For asymptotically small values of $C$, case {\it(i)}, or large detunings $|\Delta|\gg\nu,\kappa$, case {\it (ii)}, the rates assume the form of
\begin{equation}
{A'_\pm}^{\!\!\ll} = |\epsilon|^2\widetilde\eta^2 
\frac{\gamma g^2\nu^2}
{\nu^2\frac{\gamma^2}{4} + \left[\frac{\Omega_\R{L}^2}{4}-\nu(\nu\mp \delta_1)\right]^2},
\label{eq:ApmasymEIT}
\end{equation}
whose functional dependence on the parameters is the same as in EIT cooling \cite{art:morigi2003}. For analyzing case {\it (iii)} we use the definition of $C$, Eq.~\eqref{eq:singleC}, yielding 
\begin{equation}
{A''_\pm}^{\!\ll} = |\epsilon|^2\widetilde\eta^2
\frac{\gamma g^2\nu^2}
{\nu^2\frac{\gamma^2}{4} + \left[\frac{\Omega_\R{L}^2}{4}+\frac{\nu g^2\cos^2\varphi}{\nu\mp\Delta}-\nu(\nu\mp \delta_1)\right]^2}
\label{eq:ApmasymEITmod}
\end{equation}
For $\Delta=0$ it corresponds to EIT cooling with a modified Rabi-frequency $\Omega_{\rm L}^2/4\to\Omega_{\rm L}^2/4 + g^2\cos^2\varphi$. 

All cases here discussed correspond to a mean phonon number at steady state 
\begin{equation}
\langle m\rangle_{\rm st} =\frac{[\Omega^2/4-\nu(\nu+\delta_1)]^2+\gamma^2\nu^2/4}{4\nu \delta_1( \Omega^2/4-\nu^2)}\,,
\end{equation}
where $\Omega$ denotes the corresponding (modified) Rabi frequency. It achieves the minimum value, $\langle m\rangle_{\rm st,min} =(\gamma/4\delta_1)^2$, when $\Omega^2/4=\nu(\nu+\delta_1)$. The corresponding cooling rate at this value reads $W=A_--A_+\sim |\epsilon|^2 \widetilde\eta^2 4 g^2/\gamma$. 

Resonant driving, $\Delta=0$, also leads to EIT-like cooling, even for arbitrary values of the cooperativity $C$, what is especially interesting for ${\mathcal C_0}=\left.{\mathcal C}_\pm\right|_{\Delta=0}\gg 1$. Then, the transition rates between the vibrational levels are given by
\begin{align}
A_\pm^{0} = |\epsilon|^2\widetilde\eta^2 \frac{\gamma' g^2\nu^2 }{\nu^2\frac{\gamma'^2}{4}+\left[\frac{\Omega_{\rm L}^2}{4}-\nu(\nu\mp\delta_1)+\frac{\nu^2}{\kappa}\frac{\gamma'}{2}\right]^2}
\label{eq:ApmD0}
\end{align}
with the modified Rabi-frequency $\Omega_{\rm L}^2/4\to\Omega_{\rm L}^2/4+\gamma'\nu^2/2\kappa$ and the modified atomic linewidth $\gamma' = \gamma {\mathcal C}_0$. For asymptotically large ${\mathcal C_0}$ this case leads to the minimal mean vibrational occupation number at steady state of $\langle m\rangle_{\rm str, min}=(\kappa/2\nu)^2$, which can be very small provided that that the cavity linewidth is much smaller than the trap frequency.

\subsubsection{Cavity EIT cooling in the strong coupling regime}
\label{sec:suphd}

We now consider the case of large cooperativities, $C\gg 1$, and small cavity linewidths, such that $\kappa\ll\nu$. Moreover, we assume that the system is driven at three-photon resonance, $\delta_{\rm TP}=0$.

\begin{figure}
 \includegraphics[width=6.5cm]{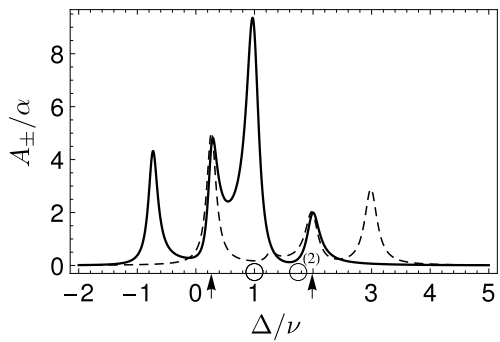}
 \caption{\label{fig:Apmsideband} Transition rates of cooling $A_{-}$ (solid line) and heating $A_{+}$ (dashed line), for large cooperativity $C\gg 1$ and small cavity decay rate,  $\kappa\ll\nu$. The arrows and circles mark the dressed states' frequencies and the dark resonances, in correspondence with Fig.~\ref{fig:dstates}. The parameters are $\gamma=10\nu$, $\kappa=\nu/10$, $g=10\nu$, $\Omega_{\rm L}=12\nu$, $\delta_1 = \delta_{\rm opt}=47.5\nu$, $\deltac=\delta_1-\nu$, $\varphi=\pi/3$, $\phi_\R{L}=\phi_\R{C}=0$. The transitions rates are given in units of $\alpha$, Eq.~\eqref{eq:alpha}, and $\Delta$ in units of $\nu$.
}
\end{figure}

When the pump is tuned on the blue sideband transition of the cavity, $\Delta=\nu$, one encounters the situation ${\mathcal C}_+=C$ and ${\mathcal C}_-\approx C\frac{\kappa^2}{4\nu^2}$, fulfilling ${\mathcal C}_+\gg{\mathcal C}_-$. The resulting transition rates between the vibrational levels take on the form
\begin{align}
A_+ &\sim |\epsilon|^2\widetilde\eta^2 \frac{1}{C}\frac{4 g^2}{\gamma}\,,\label{eq:AEITsbp}\\
A_- &= |\epsilon|^2\widetilde\eta^2 \frac{g^2\gamma(1+{\mathcal C}_-) }{\frac{\gamma^2}4(1+{\mathcal C}_-)^2+(\delta_1 - \delta_{\rm opt})^2}\,,
\end{align}
where
$$\delta_{\rm opt} = \Omega_{\rm L}^2/(4\nu)+g^2\cos^2\varphi/(2\nu)-\nu\,.$$ This corresponds to the case where the weak pump is in resonance with the red sideband of the dressed state $\ket +$ of $H_{\rm opt}$. It becomes clear from Eq.~\eqref{eq:AEITsbp} that the cooling is larger than the heating rate by the factor 
\[
\frac{A_-}{A_+}\sim\left(C^{-1}+\tfrac{\kappa^2}{4\nu^2}\right)^{-1}\,.
\] 
Additionally, diffusive processes are suppressed due to three-photon resonance. In this regime, the mean occupation number reads
\begin{align}
\langle m\rangle_{\rm st} = \left[ C\frac{1+{\mathcal C}_-}{( 1+{\mathcal C}_-)^2 + 4 (\delta_1-\delta_{\rm opt})^2/\gamma^2}-1\right]^{-1}\,,
\end{align}
and it takes on the minimal value $\langle m\rangle_{\rm st, min}\approx 1/C$ for $\delta_1 = \delta_{\rm opt}$ and ${\mathcal C}_-\ll 1$. In this limit the cooling rate reads $W\sim A_-=|\epsilon|^2\widetilde\eta^2 4 g^2/\gamma$. Figure~\ref{fig:Apmsideband} displays the transition rates $A_\pm$ for $\delta_1 = \delta_{\rm opt}$. 
For this choice of $\delta_1$ the rate $A_-$ has a pronounced maximum at $\Delta = \nu$. This maximum is displaced from the resonance around $\Delta \approx 2\nu$ by the the trap frequency $\nu$, and corresponds to the case where the pump laser is resonant with the red sideband of a dressed state. The transition rate of heating is strongly suppressed at $\Delta=\nu$ due to three-photon resonance.

\subsection{Double dark resonances}

A peculiar characteristic of this specific setup is found when the mechanical effects are predominantly governed by the interaction with the cavity field. In this case the transition rates read 
\begin{align}
 A_{\pm} \approx A_{C,\pm}=  \gamma_2 |\widetilde{\cT}_{\rm C}^{\gamma,\pm}|^2 + 2\kappa |\widetilde{\cT}_{\rm C}^{\kappa,\pm}|^2
 \label{eq:ACpm}
\end{align}
with the amplitudes $\widetilde\cT_{\rm C}^{\gamma,\pm}$, $\widetilde\cT_{\rm C}^{\kappa,\pm}$, given in Eqs.~\eqref{eq:tTC}.  These scattering amplitudes are connected to products of the functions $\cF_\gamma$ and $\cF_\kappa$, Eqs.~\eqref{eq:cF}, which determine the excitation spectra of the atom and cavity. 

\begin{figure}
\centerline{\includegraphics[width=6.5cm]{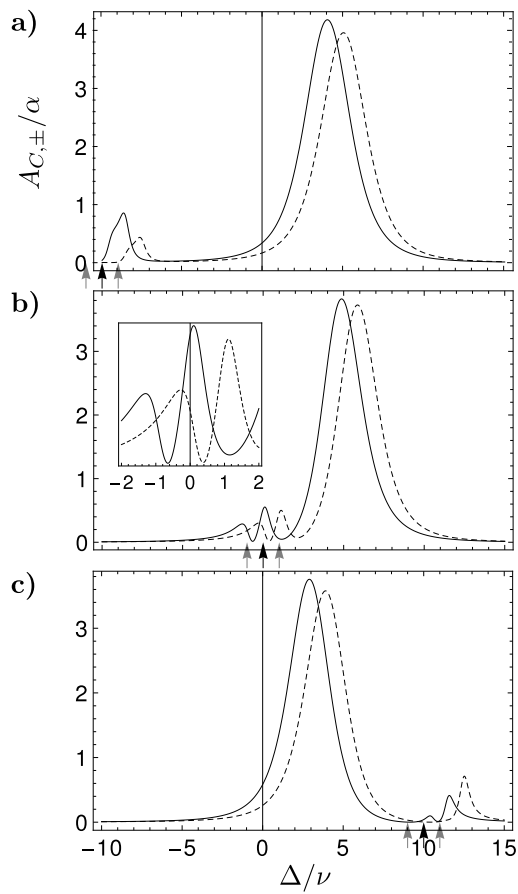}}
\caption{\label{fig:apm} 
Transition rates of cooling $A_{\rm C,-}$ (solid line) and heating $A_{\rm C,+}$ (dashed line) in units of $\alpha$, Eq.~\eqref{eq:alpha}, as a function of $\Delta$ in units of $\nu$. The rates are evaluated from Eq.~\eqref{eq:ACpm} for a) $\delta_1 = \deltac/2$, b) $\delta_1 = \deltac$ and c) $\delta_1=3\deltac/2$.  The other parameters are $\gamma=10\nu$, $\kappa=2\nu$, $g=20\nu$, $\Omega_{\rm L}=12\nu$, $\deltac=20\nu$, $\Wgeo{2}=1$, $\varphi=\pi/3$, $\phi_\R{C}=0$. The small arrows in the plots mark the positions of three-photon resonance (black) and of the corresponding sidebands, shifted by $\pm \nu$ (gray).}
\end{figure}

Figure~\ref{fig:apm} displays the rates $A_{C,\pm}$ for different values of the detuning $\delta_1$ and as a function of $\Delta$.  Cooling is found whenever $A_{\rm C,-}>A_{\rm C,+}$, that is, when the black curve overtops the gray curve. This is the case on the left edge of the broad resonance around $\Delta \approx 4\nu$. More interesting is the behavior around three-photon resonance $\Delta=\delta_1-\delta_{{\rm c}2}$, marked by the black arrow. In particular in the graphs b) and c), the rates $A_{{\rm C}\pm}$ oscillate with poles at a distance of the order of the trap frequency $\nu$, These oscillations are a consequence of Fano-like profiles in the scattering cross-section of the system and are visible in the excitation spectra of atom and cavity. Here, these approximate double dark resonances lead to alternating cooling and heating regions around three-photon resonance. 

It is interesting to consider whether the parameter can be adjusted, to suppress both carrier and blue-sideband transitions, thereby increasing the cooling efficiency. An example for such an situation is the cooling scheme discussed in Sec.~\ref{sec:suphd}. This could constitute
 a more accessible realization of a cooling process, which so far has been proposed in a tripod level configuration \cite{evers:2004} and for a two-level atomic transition confined in a resonator and at the node of the laser standing wave \cite{Zippilli_PRL05}.

\subsection{Large three-photon detuning}

The case where the control laser is far detuned from the atomic transition $\ket{g_1}\leftrightarrow\ket{e}$ corresponds to a closed two-level system consisting of the atomic levels $\ket{g_2}$ and $\ket e$ (in presence of a finite decay probability into state $\ket{g_1}$ the control laser acts as repump). In our model, this case corresponds to taking large three-photon detunings $\delta_{\rm TP}$ and large values of the detuning $|\delta_1|$. In this limit the mechanical effects are due to coherent atom-cavity interaction and to spontaneous decay, while the effect of the transverse laser coupling to the atom can be neglected. The atomic system thus reduces to an effective two-level transition coupled to a cavity which is pumped. This situation differs from the one considered in Ref. \cite{Zippilli_PRL05,Zippilli_PRA05} where the two-level atom is driven by a transverse laser. The case of a trapped two-level system whose center-of-mass motion is cooled in a driven cavity shows interesting features on its own, and its dynamics will be object of future investigations.

\section{Discussion of the results}
\label{Sec:Results}

In this section we report the graphs of cooling rate and the average vibrational number at steady state, $\langle m\rangle_{\rm st}$. These quantities, defined in Eqs.~\eqref{eq:coolingrate} and \eqref{eq:meanm}, determine the velocity of the cooling process and the final temperature, and are found by evaluating the rates in Eqs. \eqref{eq:ApmT} as a function of the physical parameters. For convenience, we report the cooling rate $\Gamma$ scaled with the frequency
\begin{align}
 \alpha=\eta^2 \Omega_\R{P}^2/\nu\,.
 \label{eq:alpha}
\end{align}
The parameters we consider base in most cases on the experiment in Ref. \cite{Meschede}. Typical values for $\alpha$, which are in accordance with Eq.~\eqref{eq:condWP}, are of the order of $10^{-6}\nu$ to $10^{-4}\nu$.

We first consider the three-photon resonance condition. This is achieved by appropriately setting the frequencies of the transverse laser and of the laser pumping the cavity. In this regime we expect that cooling is dominated by the mechanical effect stemming from the laser and cavity interaction, which mutually interfere, while the diffusion processes due to spontaneous emission are suppressed. 

\begin{figure}
\centerline{\includegraphics[width=6cm]{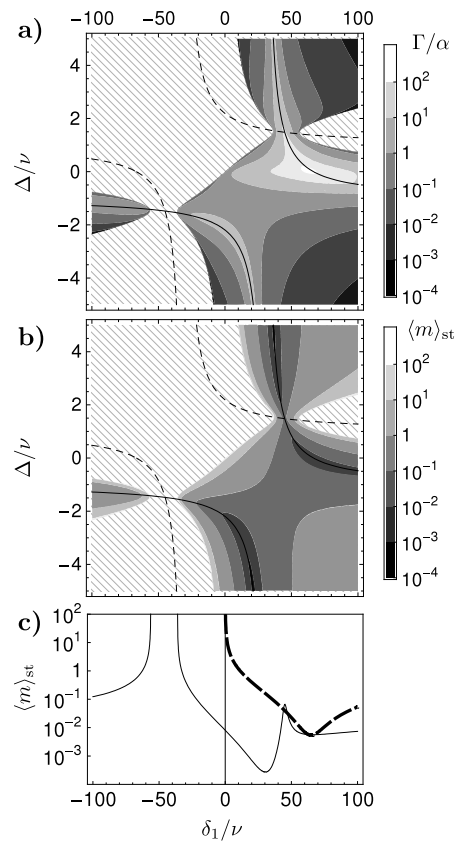}}
\caption{\label{fig:meannEIT} a) Cooling rate $\Gamma$ and b) mean phonon number $\langle m\rangle_{\rm st}$ as a function of the detunings $\delta_1$ and $\Delta$ when the three-photon resonance condition is satisfied, $\delta_{\rm TP}=0$. The solid curves in a) and b) correspond to the resonance condition from Eq.~\eqref{eq:rescondeit} which maximizes $A_-$ (the dashed line reports the curve where $A_+$ is maximum). In the hatched area the cooling rate is negative and corresponds to the region where the atomic motion is heated.  c) $\langle m\rangle_{\rm st}$ as a function of the detunings $\delta_1$ for $\Delta$ fulfilling Eq.~\eqref{eq:rescondeit} (solid line) and $\Delta=0$ (broken line). The other parameters are $\gamma=2\nu$, $\kappa=\nu/4$, $g=12\nu$, $\Omega_{\rm L}=11.2\nu$, $\varphi=\pi/3$.}
\end{figure}

Figure~\ref{fig:meannEIT} displays the contour plots of the cooling rate $\Gamma$ and of the corresponding stationary mean phonon number $\langle m\rangle_{\rm st}$ as a function of the detuning $\Delta$ between probe and cavity, and of  the detuning $\delta_1$ between control laser and the atomic transition $\ket{g_1}\leftrightarrow\ket{e}$. The solid line corresponds to the condition given in Eq.~\eqref{eq:rescondeit}, which relates $\Delta$ and $\delta_1$ and which maximizes the value of $A_-$. Indeed, the lowest values of $\langle m\rangle_{\rm st}$, and correspondingly the maxima of the cooling rate, are found along the solid line. The largest value of the cooling rate $\Gamma$, Fig. \ref{fig:meannEIT} a), is found at the intersection between this curve and $\Delta = 0$, namely, when the pump laser drives resonantly the cavity mode and the states $\ket{g_2,1}$ and $\ket{g_1,0}$ are hence also resonantly coupled. This behavior can be understood considering that the atom appears transparent to the cavity at three-photon resonance and in zero-order in the Lamb-Dicke expansion. In this regime, hence, the cavity resonance line, which scales the cooling rate, is given by a Lorentzian shape, and has its maximum when $\Delta=0$.  
Figure~\ref{fig:meannEIT}c) displays the mean phonon number as a function of $\delta_1$ when  $\Delta=0$ (solid curve). This value is compared with the one found for $\Delta(\delta_1)$, which satisfies Eq. \eqref{eq:rescondeit} and is indicated by the dashed curve. In both cases the system is driven at three-photon resonance. One observes that the temperature is minimal for the values predicted by Eq. \eqref{eq:rescondeit}.

\begin{figure}
\centerline{\includegraphics[width=6cm]{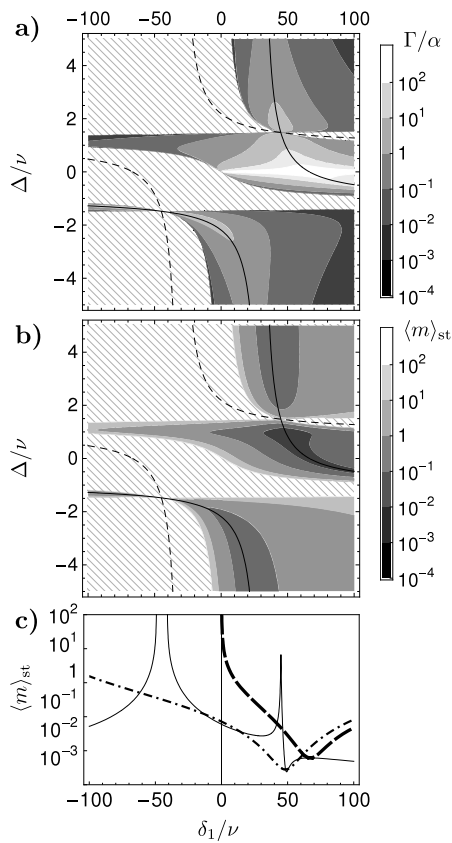}}
\caption{\label{fig:meannEITSB} Same as Fig. \ref{fig:meannEIT} but for  $\kappa\ll\nu$. The  plot c) shows cuts along the line $\Delta = 0$ (broken), $\Delta = \nu$ (dashed-dotted) and for $\Delta$ fulfilling Eq.~\eqref{eq:rescondeit} (solid line). The other parameters are $\gamma=20\nu$, $\kappa=\nu/100$, $g=12\nu$, $\Omega_{\rm L}=11.2\nu$, $\varphi=\pi/3$.}
\end{figure} 

Figure~\ref{fig:meannEITSB} displays cooling rate and mean vibrational number at steady state for the same parameters as in Fig.~\ref{fig:meannEIT}, but larger spontaneous emission rate $\gamma=20\nu$ and smaller cavity decay rate $\kappa=0.01\nu$. With respect to the results found in Fig.~\ref{fig:meannEITSB}, we observe that the cooling region about $\Delta=-\nu$ and $\delta_1<0$ becomes smaller. More striking is the appearance of a cooling region which stretches around $\Delta = \nu$ across all considered values of the detuning $\delta_1$. This is understood in terms of the cooling dynamics discussed in Sec.~\ref{sec:suphd}: In fact, for the considered parameters the rate $A_-$ is more than two orders of magnitudes larger than the rate $A_+$. In this case, the cooling transition is enhanced because the pump is tuned to the red sideband of a resonance, while the heating is suppressed because the corresponding transition coincides with a dark resonance.

Figure~\ref{fig:meannEITSB}c) displays the mean phonon number as a function of the detuning $\delta_1$ for various values of $\Delta$. The broken line corresponds to the case $\Delta = 0$ and shows a similar behavior as in Fig.~\ref{fig:meannEIT}. 
The thick dashed-dotted line is found for $\Delta=\nu$: Here, not only diffusion is suppressed due to the three-photon resonance condition, but also the heating rate is strongly reduced. Indeed, the temperature is minimal for the value of $\delta_1$ fulfilling Eq.~\eqref{eq:rescondeit} at $\Delta=\nu$.  Further insight can be gained by inspecting the dressed states shown in Fig.~\ref{fig:dstates}: One finds that the cooling regions mostly correspond with the probe laser being set on resonance with the red sideband of a narrow-line dressed state. We note, in particular, that  the condition of optimal cooling can be identified in Fig. ~\ref{fig:dstates}  at the intersection of the curve $\Delta = \delta_1-\deltac$ with the curve showing the frequency of the dressed state, shifted downwards by the trap frequency. This intersection determines the curves of optimal cooling in Fig.~\ref{fig:meannEIT} b). 

\begin{figure}
\centerline{\includegraphics[width=0.4\textwidth]{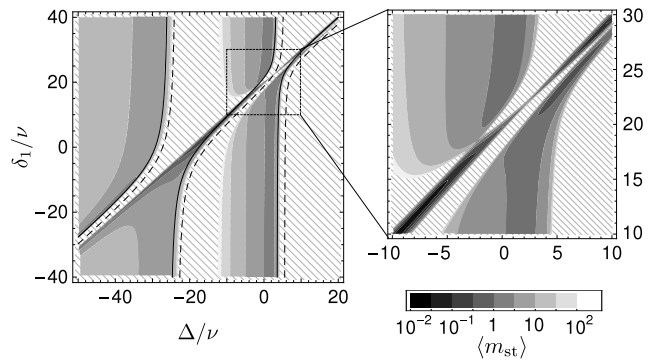}}
\caption{\label{fig:nplot} Mean phonon number $\langle m\rangle_{\rm st}$ as a function of the detunings $\Delta$ and $\delta_1$ for fixed $\deltac=20\nu$. The solid curve corresponds to  ${\rm Re}f(\Delta+\nu)=0$, Eq. \eqref{f:Delta}, which maximizes $A_-$ (dashed curve reports ${\rm Re}f(\Delta-\nu)=0$ maximizing $A_+$). The area in the box is magnified in the right plot.  The other parameters are $\gamma=10\nu$, $\kappa=2\nu$, $g=20\nu$, $\Omega_{\rm L}=12\nu$, $\deltac=20\nu$, $\Wgeo{2}=1$, $\varphi=\pi/3$.}
\end{figure}

We finally analyze the cooling dynamics when the three-photon resonance condition is not satisfied, restricting to some specific parameter regimes. Figure~\ref{fig:nplot} displays the mean phonon number at steady state as a function of $\Delta$ and $\delta_1$ at a fixed value $\deltac=20\nu$. The solid line corresponds here to the values of $\Delta$ and $\delta_1$ fulfilling Eq.\eqref{f:Delta}. For comparison, the curve corresponding to three-photon resonance is reported and corresponds to the diagonal line. Minimal temperatures are mostly found along to this curve, 
demonstrating that EIT cooling is in most parameter regimes an efficient cooling scheme.  Further parameter intervals where low temperatures are encountered are close to the curves $\Delta= -30\nu$ and  $\Delta = 0$. These regimes correspond to dynamics which can be understood in terms of Doppler cooling: the probe laser is here tuned on the red of the corresponding dressed state, whose linewidth is larger than the trap frequency. A strikingly different behavior is found close to the intersection between the three-photon resonance curve and $\Delta=0$: this parameter region is reported in the inset. Here, we observe that heating and cooling regions alternate. The cooling dynamics is here determined by interference processes: The corresponding rates $A_{\pm}$ are displayed in Fig. \eqref{fig:apm}(b), showing that the cooling rate is characterized by a double Fano-like profile.

\section{Conclusions}
\label{Sec:Conclusions}

A cooling scheme has been presented and characterized, which combines cavity quantum electrodynamics and electromagnetic induced transparency. Novel dynamics have been identified, which may allow one to efficiently cool the atomic motion. These originate from the composite action of cavity, laser, and quantized atomic motion, giving rise to interference processes which allow one to efficiently cool the motion. The cooling dynamics is robust against fluctuations of the parameters for most situations identified. Moreover, efficient ground-state cooling has been predicted for parameters which are consistent with the ones reported in Ref. \cite{Meschede}.

The theory we presented here should also be applicable to current experimental setups, where the atom is confined in traps with different potentials for the relevant atomic states, as it is often the case for a dipolar trap setup. The fluctuations of the state-dependent trapping potential may be accounted for by a modified diffusion coefficient in the rate equations, as initial considerations suggest \cite{art:morigi2003}. Moreover, the theory can be extended to the case, when the motion is only strongly confined along the cavity axis. The perpendicular motion of the atom can be taken into account by averaging $g$ over the corresponding slice of the cavity's mode function.

Future work shall address the properties of the scattered light.  For several cavity-based systems it has already been demonstrated that photons can be emitted from a single atom in a controlled way \cite{Kuhn2002, Keller2004, McKeever2004}. Furthermore, quantum interference sustains the creation of non-classical features in the emitted light \cite{Vitali,Helm2010}. Therefore, the system here presented exhibits several requirements to be potentially operating as a quantum emitter. In general, given the large number of control parameters at hand, and the possibility to interface it with several fields, this system can result to be a very useful element for a quantum network \cite{ParkinsKimble} for which further building blocks based on elementary light-atom interaction have recently been realized \cite{Boozer2007,Specht2011,Dilley2011}.

\begin{acknowledgments}
We wish to acknowledge D. Meschede and his collaborators, in particular T. Kampschulte, for many stimulating discussions which motivated this work. This work has been partially supported by the European Commission (Integrating Project AQUTE; STREP PICC), by the Ministerio de Ciencia e Innovaci{\'o}n (QOIT Consolider-Ingenio 2010; FIS2007-66944; EUROQUAM ``CMMC''), by the BMBF ``QuORep'', and by the German Research Foundation (DFG).
\end{acknowledgments}

\appendix
\section{Evaluation of the scattering rates}
\label{app:scatter}

\subsection{Calculation of the transition amplitudes for cooling}

The transition amplitudes
\begin{align}
\cT_\R{fi} = \bra{\varphi_\R{f}} \left(V+V\frac{1}{E_\R{i}-{\mathcal H}}V\right)\ket{\varphi_\R{i}}.
\label{eq:tfi}
\end{align}
of a scattering process are evaluated using the resolvent of the Hamiltonian  \cite{qob:cohentannoudji:atom_photon_interaction}. The corresponding transition rates are
\begin{align}
 R_{\rm fi} = \frac{2\pi}{\hbar}\sum_{\rm f} |\cT_\R{fi}|^2 \delta(E_\R{i}-E_\R{f})
\label{eq:srates}
\end{align}
where the $\delta$-function ensures energy conservation, and the sum goes over all relevant, different final states. The initial and final states $\ket{\varphi_\R{i}}$ and $\ket{\varphi_\R{f}}$ have defined energies $E_\R{i}$ and $E_\R{f}$ with respect to the unperturbed Hamiltonian $H_0$, Eq.~\eqref{eq:H0}. 
The total Hamiltonian ${\mathcal H} = H_0 + V + O(\eta^2)$ contains the interaction part
\begin{align}
V &= [{\mathcal H}_0-H_0] + {\mathcal H}_1 + W_{\rm P}\nonumber\\
&= W_\gamma(0) + W_\kappa + F x + \sum_{j=1,2} F_{\gamma_j} x + W_{\rm P}
\label{eq:V}
\end{align}
which includes the mechanical effects of laser and cavity $F$, Eq.\eqref{eq:force}, and the mechanical effects denoted by $F_{\gamma_j}$, Eq.~\eqref{eq:Fgammaj} of the interaction with the free radiation field in first order in $\eta$.

We regard scattering processes with initial state
\begin{equation}
\ket{\varphi_\R{i}} = \ket{\Psi_{\rm st}, m; 0_{\vec k,\epsilon}},
\label{eq:is}
\end{equation}
where the atom and cavity mode are in the state $\ket{\Psi_{\rm st}}$, Eq.~\eqref{eq:ketg20}, $m$ phononic excitations are present in the center-of-mass motion of the atom, and the free radiation field is in the vacuum state. The initial energy is $E_\R{i} = (m+\tfrac12)\hbar\nu + \hbar \Delta$. We are interested in the processes which change the  motional state of the atom by one phonon, corresponding to the final states,
\begin{subequations}
\label{eq:fss}
\begin{align}
\ket{\varphi_\R{f}^{\gamma,\pm}} = \ket{\Psi_{\rm st}, m\pm 1; 1^{(\gamma)}_{\vec k,\epsilon}}\label{eq:fsgamma}\\
\ket{\varphi_\R{f}^{\kappa,\pm}} = \ket{\Psi_{\rm st}, m\pm 1; 1^{(\kappa)}_{\vec k,\epsilon}}\label{eq:fskappa}.
\end{align}
\end{subequations}
These are states, where a photon is scattered into the modes of the electromagnetic field external to the resonator, either by spontaneous decay along $\ket e\leftrightarrow\ket{g_2}$ or cavity losses.

If one neglects the decay along transition $\ket{g_1}\leftrightarrow \ket e$, the states~\eqref{eq:fss} are the only final states for processes in first order of $\epsilon$ and $\eta_j$ where the motional state is changed by one quantum. We make the assumption $\gamma_1=0$ to avoid multi-photon scattering at the transition driven by the strong control laser. This simplification allows for a less involved analytical treatment. Comparisons with numerical results show, that the results are not significantly altered in the parameter range we are interested in.

The evaluation of the scattering amplitude $\cT_\R{fi}$, Eq.~\eqref{eq:tfi}, using the initial and final states from Eqs.~\eqref{eq:is} and \eqref{eq:fss}, is performed perturbatively in a Born series expansion
\begin{align}
\cT_\R{fi} \approx \bra{\varphi_\R{f}} \left[V G_0(E_\R{i}) V + V G_0(E_\R{i}) VG_0(E_\R{i}) V\right]\ket{\varphi_\R{i}}
\label{eq:TfiBA}
\end{align}
using the unperturbed resolvent
\begin{align}
 G_0(z) = \frac{1}{z-H_0^{\rm eff}}\,,
 \label{eq:resolvent}
\end{align}
containing the effective Hamiltonian
\begin{align}
 H_0^{\rm eff} = H_0 - i\hbar \frac{\gamma}{2} \prj{e} -i\hbar\kappa a^\dagger a\,.
\end{align}
which accounts for radiative losses.
The evaluation of Eq.~\eqref{eq:TfiBA}, using the initial and final states \eqref{eq:is} and \eqref{eq:fss}, reveals that only three processes described by the amplitudes
\begin{subequations}
\label{eq:processes}
\begin{align}
\cT_{\rm D}^\pm &= \bra{\varphi_\R{f}^{\gamma,\pm}} F_{\gamma_2}\, G_0\, W_\R{P}\ket{\varphi_\R{i}}\label{eq:cTD}\\
\cT_{\rm F}^{\gamma,\pm} &= \bra{\varphi_\R{f}^{\gamma,\pm}} W_{\gamma_2}(0)\, G_0\,F\, G_0 W_\R{P}\ket{\varphi_\R{i}}\label{eq:cTFg}\\
\cT_{\rm F}^{\kappa,\pm} &= \bra{\varphi_\R{f}^{\kappa,\pm}} W_{\kappa} \,G_0\,F\, G_0\, W_\R{P}\ket{\varphi_\R{i}}\label{eq:cTFk}
\end{align}
\end{subequations}
contribute. Here, $G_0\equiv G_0(E_\R{i})$.
Equation~\eqref{eq:cTD} represents the diffusive process due to the mechanical effect of spontaneous emission, whereas Eqs.~\eqref{eq:cTFg} and \eqref{eq:cTFk} stand for the lowest order mechanical effects of the control laser and the cavity.

With the specific definitions given in this appendix, it is easy to check that the scattering amplitudes, Eqs.~\eqref{eq:processes}, are the only contributions to light scattering including mechanical interaction up to first order in $\eta$ and $\epsilon$. They are straightforwardly evaluated by substituting the various operators by the explicit expressions from Eqs.~\eqref{eq:WP}, \eqref{eq:Wgammaj}, \eqref{eq:force}, \eqref{eq:Fgammaj}. The matrix elements of $G_0(E_\R{i})$ are evaluated in App.~\ref{app:resolvent}, in particular Eqs.~\eqref{eq:G0block} and~\eqref{eq:G0explicit} with $E_\R{i} = \hbar\Delta+\hbar  (m+\tfrac12)\nu$. One finds
\begin{align}
\cT_{\rm D}^{\pm}&=\hbar\sqrt{m+\delta_\pm} [\vec e_x\cdot\vec e_{\vec k,\epsilon}] g_{\vec k,\epsilon}^{(\gamma_2)}\widetilde{\cT}_{\rm D}(|\vec k|),\\
\cT_{\rm F}^{\gamma,\pm}&=\hbar\sqrt{m+\delta_\pm}  g_{\vec k,\epsilon}^{(\gamma_2)}\widetilde{\cT}_{\rm F}^{\gamma,\pm}(|\vec k|),\\
\cT_{\rm F}^{\kappa,\pm}&=\hbar\sqrt{m+\delta_\pm}  g_{\vec k,\epsilon}^{(\kappa)}\widetilde{\cT}_{\rm F}^{\kappa,\pm}(|\vec k|),
\end{align}
after collecting all remaining factors -- which are independent of the polarization $\vec e_{\vec k, \epsilon}$ and direction $\vec k/|\vec k|$ of the scattered photon and the vibrational quantum number $m$ -- in the tilded quantities. The abbreviation $\delta_\pm$ yields 1 for the ``$+$'' case and otherwise vanishes.

The corresponding rates are found using Eq.~\eqref{eq:srates}, where one has to sum over all relevant polarizations and wave vectors,
\begin{align*}
&R_{\rm D}^{\pm}=2\pi(m+\delta_\pm) {\sum_{\vec k,\epsilon}}^{(\gamma_2)} [\vec e_x\cdot\vec e_{\vec k,\epsilon}]g_{\vec k,\epsilon}^{(\gamma_2)}\delta( c|\vec k| - \omega_{e2})|\widetilde{\cT}_{\rm D}|^2,\\
&R_{\rm F}^{\gamma,\pm}=2\pi(m+\delta_\pm) {\sum_{\vec k,\epsilon}}^{(\gamma_2)} |g_{\vec k,\epsilon}^{(\gamma_2)}|^2\delta(c|\vec k| - \omega_{e2})|\widetilde{\cT}_{\rm F}^{\gamma,\pm}|^2,\\
&R_{\rm F}^{\kappa,\pm}=2\pi(m+\delta_\pm){\sum_{\vec k,\epsilon}}^{(\kappa)} |g_{\vec k,\epsilon}^{(\kappa)}|^2\delta( c|\vec k| - \omega_{\rm C})|\widetilde{\cT}_{\rm F}^{\kappa,\pm}|^2.
\end{align*}
Moreover, we used $\omega_{\rm P}\approx\omega_{\rm C}\approx\omega_{e2}\gg\nu$ to approximate the argument of the delta function, and $\omega_{e2}=\omega_e-\omega_2$. The quantities $\widetilde{\cT}_{\rm D}$, $\widetilde{\cT}_{\rm F}^{\gamma,\pm}$ and $\widetilde{\cT}_{\rm F}^{\kappa,\pm}$ in the latter equations are evaluated at the $|\vec k|$-value determined by the delta function. Taking into account the definitions~\cite{qob:cohentannoudji:atom_photon_interaction}
\begin{align}
\gamma_2 = \frac{2\pi}{\hbar}\sum_{\vec k,\epsilon} |\hbar g_{\vec k,\epsilon}^{(\gamma_2)}|^2\delta(\hbar c|\vec k|-\hbar\omega_{e2})\label{eq:defgamma},\\
\kappa = \frac{\pi}{\hbar} \sum_{\vec k,\epsilon} |\hbar g_{k,\epsilon}^{(\kappa)}|^2 \delta(\hbar c|\vec k|-\hbar\omega_{\rm C}),\label{eq:defkappa}
\end{align}
and $A_\pm (m+\delta_\pm) = R_{\rm D}^{\pm} + R_{\rm F}^{\gamma,\pm} + R_{\rm F}^{\kappa,\pm}$ \cite{qo:stenholm1986} leads to expression~\eqref{eq:ApmT}, where $\Wgeo{2}$ was introduced as a geometrical factor characterizing the atomic dipole radiation pattern~\cite{qo:javanainen:1980}.

\subsection{Calculation of the excitation spectra} 
\label{app:excspec}

The excitation spectra of the atom at rest give the rate of photon scattering as a function of the probe frequency $\omega_\R{P}$ either at the cavity output, giving $S_{\rm exc}^{\kappa}(\omega_\R{P})$, or from the resonance fluorescence of the atom integrated over the solid angle involving the modes external to the resonator, which we label by $S_{\rm exc}^{\gamma}(\omega_\R{P})$. It has the form $S_{\rm exc}^{j} \propto R^j(\omega_{\rm P})$, for $j=\kappa,\gamma$, whereby $R^j(\omega_{\rm P})$ is the rate of scattering a probe photon, assuming the atom-cavity system being in the initial state $|g_2,0\rangle$ at energy $E_\R{i}=\hbar\omega_\R{P}$, and that the atomic motion is neglected. They are connected to the scatter amplitudes
\begin{align}
\cT^{j}=\bra{\varphi_{\rm f}}W_j(0) G_0(E_\R{i}) W_{\rm P}\ket{\varphi_{\rm i}}.
\label{eq:processexc}
\end{align}
by Eq.~\eqref{eq:srates}, evaluated up to first order in the small parameter $\epsilon$. 
These amplitudes correspond to processes, where, starting from the state $\ket{\varphi_{\rm i}}$, Eq.~\eqref{eq:is}, the cavity is excited by the probe laser as described by $W_{\rm P}$. The subsequent evolution determined by $G_0(E_\R{i})$ couples the states of the manifold $\{\ket{e,0}, \ket{g_1,0},\ket{g_2,1}\}$  due to the strong cavity-atom and control laser interaction. Generally, all states of the manifold can be populated at this intermediate state, and a photon can leave the atom-cavity system either by spontaneous emission ($W_j = W_{\gamma_2}(0)$) or by leaking through the cavity mirrors ($W_j=W_\kappa$), leading to the final state $\ket{\varphi_{\rm f}}=\ket{g_2,0,m;1_{\vec k,\epsilon}}$. We note that for spontaneous decay it is sufficient to take into account only $W_{\gamma_2}(0)$, even for $\gamma_1\neq 0$: Both scattering amplitudes only differ by a constant prefactor. 

Evaluating Eq.~\eqref{eq:processexc} yields
\begin{align}
\cT^{\gamma} = \hbar \frac{\Omega_{\rm P}}{2}g_{\vec k,\epsilon}^{(\gamma_2)}\hbar\bra{e,0}G_{\rm opt}^{(1)}(\hbar\Delta)\ket{g_2,1},\\
\cT^{\kappa} = \hbar \frac{\Omega_{\rm P}}{2}g_{\vec k,\epsilon}^{(\kappa)}\hbar\bra{g_2,1}G_{\rm opt}^{(1)}(\hbar\Delta)\ket{g_2,1},
\end{align}
where the required matrix elements of $G_{\rm opt}^{(1)}(\hbar\Delta)$, calculated in App.~\ref{app:resolvent}, give the functions $\cos\varphi\cF_\gamma(\Delta)$, Eq.~\eqref{eq:cFgamma}, and $\cF_\kappa(\Delta)$, Eq.~\eqref{eq:cFkappa}, respectively. Using Eq.~\eqref{eq:srates} and after summing over all relevant modes $\vec k$, $\epsilon$ of the emitted photon, one finds the excitation spectra in the form of Eqs.~\eqref{eq:excspecs}.

\section{Matrix elements of the resolvent}
\label{app:resolvent}

In this appendix the relevant matrix elements of the resolvent $G_0(z)$, Eq.~\eqref{eq:resolvent}, are calculated.
We first rewrite
\begin{equation}
G_0(z) = G_{\rm opt}(z-H_{\rm ext}-H_{\rm free})
\label{eq:G0block}
\end{equation}
using
$G_{\rm opt}(z) = [z-H^{\rm eff}_{\rm opt}]^{-1}$ and
\begin{align}
H^{\rm eff}_{\rm opt} = H_{\rm opt} - i\hbar\frac{\gamma}2\prj{e} - i\hbar\kappa a^\dagger a\,.
\label{eq:Hopteff}
\end{align}
The operator $G_{\rm opt}(z)$
has block diagonal form ($P_n H^{\rm eff}_{\rm opt}P_m = 0$ for $n\neq m$) when considering the subspaces of the state  $\{\ket{g_2,0}\}$ and the manifolds
\begin{align}
\cM_{n} = \{\ket{g_2,n},\ket{e,n-1},\ket{g_1,n-1}\}\quad (n>0)\,,
\label{eq:manifold}
\end{align}
where we denoted the corresponding projectors by  $P_0=\prj{g_2,0}$ and $P_n$, respectively.

For the dynamics analyzed in this work it is sufficient to calculate the matrix elements belonging to the case $n=1$. The relevant matrix elements are found by inverting $z-P_1 H_{\rm opt}^{\rm eff} P_1$, yielding
\begin{widetext}
\begin{align}
G_{\rm opt}^{(1)}(\hbar \zeta) = \frac{1}{\hbar f(\zeta)}
\Big[ &\prj{e,0}\;\left\{(\deltac-\delta_1+\zeta)(i\kappa+\zeta)\right\}\nonumber\\
+&\prj{g_2,1}\left\{\left(\deltac+i\frac{\gamma}2+\zeta\right)(\deltac-\delta_1+\zeta)-\frac{\Omega_{\rm L}^2}4\right\}\nonumber\\
+&\prj{g_1,0}\left\{\left(\deltac+i\frac{\gamma}2+\zeta\right)(i\kappa+\zeta)-g^2\cos^2\varphi\right\}\nonumber\\
+&g\cos\varphi\frac{\Omega_{\rm L}}{2}\left[\ketbra{g_1,0}{g_2,1}+{\rm H.c.}\right]\nonumber\\
+&(i\kappa+\zeta)\frac{\Omega_{\rm L}}{2}\left[\ketbra{g_1,0}{e,0}+{\rm H.c.}\right]\nonumber\\
+&g\cos\varphi(\deltac-\delta_1+\zeta)\left[\ketbra{g_2,1}{e,0}+{\rm H.c.}\right]\Big].
\label{eq:G0explicit}
\end{align}
\end{widetext}
All the matrix elements of Eq. \eqref{eq:G0explicit} are inversely proportional to the function
\begin{align}
f(\zeta) =& \det[\zeta-P_1 H^{\rm eff}_{\rm opt}P_1/\hbar]
= \prod_{\ell=\pm,\circ}(\zeta-\omega^{\rm eff}_{\ell})\nonumber\\
=& (i\kappa+\zeta)\left\{\left[\deltac+\zeta+i\frac{\gamma}2\right](\deltac+\zeta-\delta_1)-\frac{\Omega_{\rm L}^2}{4}\right\}\nonumber\\
&-g^2\cos^2\varphi(\deltac+\zeta-\delta_1).
\label{eq:fz}
\end{align}
Its poles are the complex eigenfrequencies $\omega_\pm^{\rm eff}$, $\omega_\circ^{\rm eff}$ of the non-Hermitian operator $H^{\rm eff}_{\rm opt}$, Eq.~\eqref{eq:Hopteff}, reduced to the states within the manifold $\cM_1$. The real and imaginary part of these eigenfrequencies are plotted in Fig.~\ref{fig:dstates} and determine the relevant resonances of the scattering processes.

%

\end{document}